# Lipidomic approach for stratification of Acute Myeloid Leukemia patients

Christian Thiede[2], Gerhard Ehninger[2], Kai Simons[1,3], Michal Grzybek[4,5], Adam Stefanko[1,*]

[1] Max Planck Institute of Molecular Cell Biology and Genetics, Dresden, Germany

[2] Medical Clinic and Polyclinic I, University hospital TU Dresden, Dresden, Germany

[3] Lipotype GmbH, Dresden, Germany

[4] Paul Langerhans Institute Dresden of the Helmholtz Centre Munich at the University Clinic Carl Gustav Carus, TU Dresden, Dresden, Germany

[5] German Center for Diabetes Research (DZD e.V.), Neuherberg, Germany

* corresponding author

**Corresponding author:**

Dr. Adam Stefanko

Max Planck Institute MPI-CBG, Dresden, Germany

Pfothenhauerstr. 108

01307 Dresden

stefanko@mpi-cbg.de

**Abstract**

The pathogenesis and progression of many tumors, including hematologic malignancies is highly dependent on enhanced lipogenesis. De novo fatty-acid synthesis permits accelerated proliferation of tumor cells by providing structural components to build the membranes. It may also lead to alterations of physicochemical properties of the formed membranes, which can have an impact on signaling or even increase resistance to drugs in cancer cells. Cancer type-specific lipid profiles would allow understanding the actual effects of lipid changes and therefore could potentially serve as fingerprints for individual tumors and be explored as diagnostic markers. We have used shotgun MS approach to identify lipid patterns in different types of acute myeloid leukemia (AML) patients that either show no karyotype changes or belong to t(8;21) or inv16 types. The observed differences in lipidomes of t(8;21) and inv(16) patients, as compared to AML patients without karyotype changes, concentrate mostly on substantial modulation of ceramides/sphingolipids synthesis. Also significant changes in the physicochemical properties of the membranes, between the t(8;21) and the other patients, were noted that were related to a marked alteration in the saturation levels of lipids. The revealed differences in lipid profiles of various AML types increase our understanding of the affected biochemical pathways and can potentially serve as diagnostic tools.

**Introduction**

Acute Myeloid Leukemia (AML) is a clonal bone marrow disorder resulting from diverse phenotypic and genetic alterations in the differentiation of hematopoietic stem cells causing excessive proliferation, growth and accumulation of abnormal immature leukemic neoplastic cells, called blasts[1-4]. The major causes of AML are deregulations in one or more of the numerous components of the signaling networks that control cell growth and proliferation, either by a gain-of-function mutations or their overexpression[5,6]. Most efforts towards molecular characterization of AML have been directed towards changes in the genome, transcriptome or proteome [7-12], while it becomes more evident that pathogenesis and progression of cancer not only involves changes in cellular lipidomes but might even be accelerated by these[13-19]. In fact, the hallmark of many tumors, including hematologic malignancies is enhanced lipogenesis, arising from increased activities of fatty acid biosynthetic enzymes (Acc1, Fasn and Scd1)[20-23]. Accelerated fatty acid synthesis may permit faster growth of tumor cells. Furthermore, synthesis of monounsaturated fatty acids may have additional benefits as it has been shown that oleate can protect cells against saturated fatty acid toxicity and cellular stress[18,22,24,25]. Acc1, Fasn and Scd1 have therefore been identified as plausible targets for cancer therapy[14,18,20-23]. One of the lipid classes, which has been associated with cancer pathogenesis are sphingolipid metabolites like ceramides, lactosylceramide and sphingosine-1-phosphate, which act as effector molecules[15]. Ceramide in particular is intimately involved in the regulation of cancer-cell growth, differentiation, senescence and apoptosis[26-28]. Attenuation of ceramide levels by its conversion to glucosylceramide contributes to the multidrug resistance of several human cancer types, including leukemia, breast cancer, melanoma and neuroblastoma[17,29-31]. Moreover, sphingolipids are a family of membrane lipids with important structural roles in the regulation of the fluidity and subdomain structure of the lipid bilayer, especially lipid rafts[32]. These nanodomains act as regulatory platforms for a number of growth factor receptors e.g. EGFR, c-kit, VEGFR etc. controlling the initiation of signaling cascades that regulate cell growth and differentiation[33,34]. Therefore, deregulation of lipid metabolism might in consequence lead to perturbations in domain structures within the cellular membranes. Until today regulatory mechanisms that govern lipid metabolism and homeostasis are largely unknown. The metabolic network of enzymes and factors that regulates the enormous heterogeneity of the distinct lipid species is poorly understood. Notably, metabolic transitions that affect lipid metabolism are prone to induce lipidome-wide ripple effects and prompt compensatory metabolic responses[35,36]. Therefore there is a growing interest in understanding the effects of lipid changes in order to uncover lipid profiles that could potentially serve as a unique cellular fingerprint for individuals and be explored as diagnostic markers.

The purpose of this study was to explore whether lipidomics can be integrated into clinical research by characterizing the lipidomic profiles of AML patients. We have focused on three subsets of patients with specific clinical and biological AML characteristics: translocation t(8;21), inversion inv16 and patients with normal karyotype (AML-nk). Chromosomal abnormalities are detected in approximately 60% of newly diagnosed AML patients among which the most common are t(8;21) and inv16 that affect the expression of the Core Binding Factor (CBF), a heterodimeric transcription factor controlling key programs in cell survival[37-39], at least in part, through regulating sphingolipid metabolism [40,41]. The t(8;21) and inv16 are usually associated with good prognosis and survival. The 'normal karyotype' AML group is very heterogeneous as it emanates from a number of factors that consist of aberrant signal transduction pathways and cause uncontrolled growth and inhibition of apoptosis[2,6,42]. Moreover, these changes are often associated also with increased activity of multidrug resistance proteins and renewal of leukemia stem cells[30,31].

We applied shotgun mass spectrometry approach to analyze the cellular lipidomes of the various AML types and we could identify marked differences among the various AML types. The observed lipid patterns can therefore be used to identify metabolic pathways targeted in personalized anti-cancer therapies.

## Materials and Methods

### Leukocytes collection

t(8;21), inv16 and normal karyotype (AML-nk) AML samples were collected from patients at the time of diagnosis, so before any AML specific therapy was applied. The samples, and were stored as cryopreserved stocks. Full description of patients can be found in **Table1**.

### Lipids and lipid standards

Synthetic lipid standards: cholesterol (Chol-d6) (deuterated), sterol ester (CE 20:0), triacylglycerol (TAG 17:0-17:0-17:0), diacylglycerols (DAG 17:0-17:0), phosphatidic acid (PA 17:0-17:0), phosphatidylcholine (PC 17:0-17:0), phosphatidylethanolamine (PE 17:0-17:0), phosphatidylglycerol (PG 17:0-17:0), phosphatidylserines (PS 17:0-17:0), phosphatidylinositol (PI 16:0-16:0), lysoglycerophospholipids (LPA 17:0, LPC 12:0, LPE 17:1, LPS 17:1, LPI 17:1), sphingomyelin (SM 18:1;2-12:0) and ceramides (Cer 18:1;2-17:0, GlcCer 18:1;2-12:0, LacCer 18:1;2-12:0) were purchased from Avanti Polar Lipids. Total protein concentration of leukemic cells was estimated using the microBCA Protein Assay Kit (Thermo Scientific). Aliquots of cells equivalent to 20-25ug of protein that correspond to ~250 000 cells, were transferred into 2ml tubes (Eppendorf AG), pelleted (1000g, 5') washed twice in 1000µl of 155mM ammonium bicarbonate and were spiked with 10µl mixture of internal lipid standards dissolved in methanol. All samples were subjected to two-step lipid extraction with 750µl of 10:1 chloroform:methanol (C/M) (v/v) followed by 2:1 C/M mixture at 4°C, respectively. The lipid extracts were collected from organic phases and evaporated under vacuum. Dry lipid extracts were re-dissolved in 100µl of chloroform/methanol 1:2 (v/v) and analyzed on Q-Exactive (Thermo Fisher Scientific) instrument equipped with electrospray ion source - ESI TriVersa Nanomate (Advion Biosciences). Three independent runs were done for each lipid extract.

**Shotgun Lipidomics Analysis and Data Processing**

Lipid extracts (7.5µl) from the first extraction step were diluted in 13mM ammonium acetate in propanol (10µl) to achieve the final solvent composition of 7.5mM ammonium acetate in chloroform/methanol/propanol 1:2:4 (v/v). Lipid extracts from the second extraction step were dissolved in 10µl of 0.01% methylamine as an ion modifier and subjected to negative mode MS analysis. Lipid extracts from 10:1 organic phase were analyzed with polarity switching in both positive and negative ion mode, the 2:1 extracts in negative only. Positive ion mode analysis was performed using FTMS with target resolution of 280,000 at m/z 200 and MS/MS with 17,500 at m/z 200 to monitor CE, DAG and TAG species. Negative FTMS mode was chosen to monitor lyso-phospholipids (LPG, LPS, LPI, LPS, LPC, LPE) and combined with tandem MS/MS for selected phospholipids (PC, PC O-, PE, PE O-, PS, PA, PI, PG), sphingolipids (SM, Cer, Hex-Cer, diHex-Cer) globosides (Gb3 and Gb4) and ganglioside GM3. The same, targeted resolution settings were used for both polarities. Cholesterol was measured separately after chemical acetylation[43]. Briefly, 20µl of 10:1 lipid extract was dried down and 75µl of acetyl chloride in 1:2 C/M (v/v) added. After an hour incubation samples were dried down, re-dissolved in 50µl C/M 1:2 and analyzed in positive ion mode, using similar settings as described above. Molar amounts of lipid species were determined using spiked-in internal quantitative standards.

Samples were infused into the Q-Exactive instrument with TriVersa NanoMate ESI source. Data acquisition was performed in both positive and negative mode with polarity switching where MS was used for quantification and MS/MS for fatty acids and lipid precursors identification. Data were analyzed and de-convoluted by LipidXplorer software. Data visualization and normalization (pmol, mol%) calculations were performed in Microsoft Excel, GraphPad Prism 6.0 and SIMCA 14.0 software.

**GP measurements**

For fluidity analysis, dried lipids were rehydrated in 150 mM NaCl, 25 mM HEPES pH 7.25. The resulting liposomes were subsequently subjected to 10 freezing/thawing cycles (liquid nitrogen/37°C) and extruded through 100nm polycarbonate filters. Finally, vesicles were stained (15 min) with 2 µM C-laurdan. Fluorescence emission spectra were recorded (Ex. 385 nm, Em. 400–550 nm) and analyzed as described in Kaiser et al.[44]

## RESULTS

**Lipidome analysis reveals significant differences among the AML subtypes**

Total leukocyte fractions containing 250 000 cells, corresponding to approximately 20-25µg of total protein content, were subjected to two-step lipid extraction and used for MS lipidomics measurements and GP measurements. Shotgun lipidomics MS and MS/MS designed acquisition routine together with in-house developed lipidomics software enabled to profile more than 400 unique lipid species of the 25 analyzed lipid classes.

To gain a better understanding of the observed variations among the lipidomes, multivariate statistical data analysis was applied. For dimension reduction and visualization of the data, we have performed Principal Component Analysis with Pareto scaling in SIMCA software. This statistical tool is used to emphasize variation and bring out strong patterns in a dataset. PCA was applied to lipidomics data that were first normalized to the internal standards and transformed to mol% values. Subsequently two-dimensional score plot was generated (R2X values were 0.253 and 0.094 for the first and second principal components, respectively) (Figure 1), which showed that t(8;21) samples were clearly separated from the inv16 and AML-nk patient samples, that were mixed within one cluster. This suggested that we should expect substantial differences between the lipidomes of t(8;21) and the other AML types.

To check if the lipidome differences are significant we have compared the lipidomes of different AML types among each other using the orthogonal partial least square- discriminant analysis. In all cases, the differences appeared significant ($p<0.01$), also in the case of inv16 and AML-nk that formed a mixed cluster in the PCA plot. Next, the actual differences in lipid species were analyzed by comparing the various AML types among each other. Briefly, lipid fold change followed by multiple t tests analysis was calculated, which allowed the identification of unique species with the most pronounced difference (Figure 2). The major difference between the t(8;21) and AML-nk was observed for ceramide backbone-containing lipids: SM, Cer, Hex-Cer and GM3 ganglioside (Figure 2A). Decrease of SM in t(8;21) was correlated with an increase in the levels of ceramides (up to 3-fold), Hex-Cer (up to 5-fold) and GM3 (up to 15-fold) suggesting a shift towards synthesis of glycosphingolipids. Unfortunately it is not clear if GM3 derivatives, bearing additional sugar modifications, are also upregulated since these cannot be currently measured quantitatively with shotgun MS technique. When we compared the t(8;21) and inv16 groups to each other (Figure 2B), we also observed mostly changes in the expression profiles of lipids belonging to the sphingolipid pathways. We observed a decrease in the levels of SM and an increase in the synthesis of the glycosphingolipids (Figure 2C). Interestingly the t(8;21) seem also to contain more ceramides than inv16, suggesting that this type of leukemic cells might be more prone to apoptosis[26-28]. The inv16 and AML-nk patients represent a mixed population in the PCA plot (Figure 1). Nevertheless, direct comparison of these two groups also points to noticeable lipidome differences ($p<0.001$) (Figure 2c). The most prominent change was the increased abundance of various species of Hex-Cer in the inv16 samples. These lipids are the precursors for production of other glycolipids, but they have been also suggested to play an important role in multidrug resistance[17,29,31]. It seems like we observed a specific accumulation of this particular lipid class, rather than a shift in glycosphingolipid metabolism, as observed for the t(8;21) cells.

**Measurement of Generalized Polarization Index reveal dramatic changes in membrane fluidity in t(8;21) cells.**

Activation of the lipogenic pathway at early stages of cancer pathogenesis is critical for proliferation of various types of tumors including hematologic malignancies[21,23]. The majority of newly synthesized fatty acids in cancer cells are converted predominantly to membrane phospholipids. Specifically, the proportion of monounsaturated and polyunsaturated fatty acids in the major phospholipids was reported to be significantly higher in cancer compared to noncancerous tissues[45]. Here we compared each type of leukemia to see if there are important differences in saturation levels of lipids. A marked difference was observed, especially in case of the t(8;21) leukemia, where a significant shift towards polyunsaturated fatty acids at the cost of monounsaturated was noted (Figure 3A). This was however, clearly observed only for the membrane lipids (phospholipids+sphingolipids), which consist roughly 80% of all the measured lipids, but not for the group of storage lipids (TAG, DAG, SE) (Supplementary Figure 1). Alterations in fatty acid saturation can dramatically influence membrane properties and in consequence affect many aspects of the cellular machinery. Higher level of unsaturation may result in higher cell growth and proliferation or protection from ER stress. On the other hand it has been reported that the shift towards increased levels of saturated and monounsaturated phospholipids, potentially protects cancer cells from oxidative damage by reducing lipid peroxidation[14,18].

Fatty acid saturation and length or cholesterol content are the 3 major components influencing membrane fluidity. Since we observed a significant shift of fatty acid saturation and we did not observe major changes in cholesterol abundance or fatty acid length (Figure 3 and Supplementary Figure 1) we speculated that there should be substantial change in membrane fluidity between t(8;21) and the other AML types. To validate our data, we have prepared small unilamellar vesicles from the extracted lipids and we measured membrane fluidity (General Polarization Index, GP) using c-laurdan as described elsewehere[46]. Indeed, a tight correlation between the measured levels of unsaturation and GP values was obtained, confirming that t(8;21) samples contain more fluid membranes. These findings underline the importance of precise measurements of lipid and desaturation levels in cancer cells in order to understand which metabolic pathways are aberrantly regulated and therefore could be potential targets for therapy.

**Discussion**

Similar to embryonic cells, cancer cells are highly dependent on de novo lipogenesis for their proliferation, and the lipogenic pathway is activated at a relatively early stage in various types of tumors[18,21,47]. Moreover, there seems to be a correlation between fatty acid synthase expression, which is highly increased in metastatic tumors, increased membrane fluidity and the decreased patient survival[48-51]. The majority of newly synthesized fatty acids in cancer cells are converted predominantly to phospholipids, which are then incorporated into membranes of proliferating cancer cells[18]. Changes in lipid metabolism disturbing equilibrium among the saturation levels of fatty acids incorporated in the membrane lipids might lead to changes in membrane fluidity and affect the structure and composition of membrane domains (rafts) from which many of the important signaling cascades originate. In consequence aberrant signal transduction might enhance the proliferation and survival of hematopoietic progenitor cells[2,6]. Additional layer of complexity arises as a consequence of the metabolic interconversions of one lipid to another — e.g. conversion of PI(4,5)P2 to DAG or DAG/PC to PA, which again may have pronounced effect on signaling events[52,53]. Similarly many sphingolipid-regulated functions have significant and specific links to various aspects of cancer initiation, progression and response to anticancer treatments e.g. hydrolysis of ceramide produces sphingosine, the phosphorylation of which yields sphingosine-1-phosphate that regulates cell growth and suppresses programmed cell death[54-56]. As such, the many pathways of sphingolipid metabolism constitute an interconnected network that not only regulates the levels of individual biologically active molecules, but also their interconversion and, ultimately, the balance among them. Thus the difference in lipid metabolism between tumors and normal tissues renders the lipogenic pathway an attractive target for anti-cancer therapies, which so far was exploited moderately, at best[23,57].

It should be noted that in both t(8;21) and inv16 AML types, due to karyotype changes, deregulation of CBF expression occurs and it has been reported previously that CBF controls the expression of some of the genes involved in sphingolipid metabolism[40,41]. Its overexpression has been shown to reduce intracellular long-chain ceramides in NIH3T3 fibroblasts and to elevate extracellular sphingosine-1-phosphate levels[40]. On the other hand, in CBFβ-silenced cells substantial increases in ceramide species and decreases in lactosylceramides were identified that correlated with enhanced cell death[41].

Here for the first time we show and quantify the changes in the lipidomes of these cells. Indeed the major differences discovered between the various AML types consist of changes in the sphingolipids and ceramides. Interestingly, although both t(8;21) and inv16 share similarities in genetic background (deregulation of CBF expression), there are still significant differences between their sphingolipid metabolism. The t(8;21), as compared to inv16 and AML-nk, shows substantial decrease of sphingomyelin, probably a result of decreased expression of SGMS1 and SGMS2 genes[7], which detours ceramides to glycosphingolipids synthesis pathway. In inv16 samples the sphingolipid metabolism looks much more equilibrated and only a specific accumulation of hexosylceramides is observed (Figure 2C), which has been reported as a multidrug resistance phenotype in many cancer cells[17,29-31]. The observed pattern is particularly interesting as in t(8;21) and inv16 patients sensitive to ABT737 (Bcl-2 inhibitor) and BV6 (antagonist of inhibitor of apoptosis proteins) drugs, genes responsible for ceramide synthesis were upregulated in contrast to patients resistant to these drugs[7]. Changes in expression of enzymes of sphingolipid metabolism in several cancers are consistent with the emerging tumor-suppressor functions of ceramide. Our data provide evidence that ceramide levels significantly differ among various patients, even those bearing similar genetic perturbations, which might directly correlate with susceptibility of these patients to chemotherapy.

Among all studied groups, the t(8;21) patients display highest membrane fluidity, which must be the effect of an increased unsaturation level within the fatty acid moieties of the membrane lipids. Increased membrane fluidity has been associated with increased cell growth and proliferation in many cancers[58-60]. It has been also previously shown to modulate certain receptor tyrosine kinase proteins, resulting in disturbed signaling[61-64]. In line with this, some drugs are membrane active, and it has been suggested that the ability of a drug to alter a membrane fluidity correlates with the drug activity[65,66].

AML with its mosaic etiological factors necessitates a systemic approach for mechanistic understanding and for optimization of therapeutical targets. Lipid signatures quantified here correlated with cytogenetic abnormalities of leukemic cells, which paves the way towards new diagnostics and could offer a prognostic value for clinical patient evaluation.

## Conclusion

Mass spectrometry-driven quantitative analysis of cellular lipids, especially the so-called shotgun lipidomics technique is increasingly being explored for its potential as a clinical diagnostic tool. A major benefit of this technology is that hundreds of lipid species can be directly identified and accurately quantified in a relatively short analysis time[67-69]. This work demonstrates that a lipidomic analysis of acute myeloid leukemia cells could be employed for stratification of hematological malignancies. The observed imbalances of lipid synthesis pathways might directly contribute to disease progression.

**Acknowledgements**
We thank Christian Klose and Patrycja Dubielecka for critical discussions and reading of the manuscript. We acknowledge that the results of this research have been achieved using the German Federal Ministry of Education and Research grant to the German Center for Diabetes Research (DZD e.V.) (M.G.).

**Author contribution**
AS and MG performed experiments. CT and GE contributed samples. AS, CT, GE, KS and MG analyzed data. AS, KS and MG wrote the manuscript.

**Conflict of interests**
The authors declare no conflict of interests.

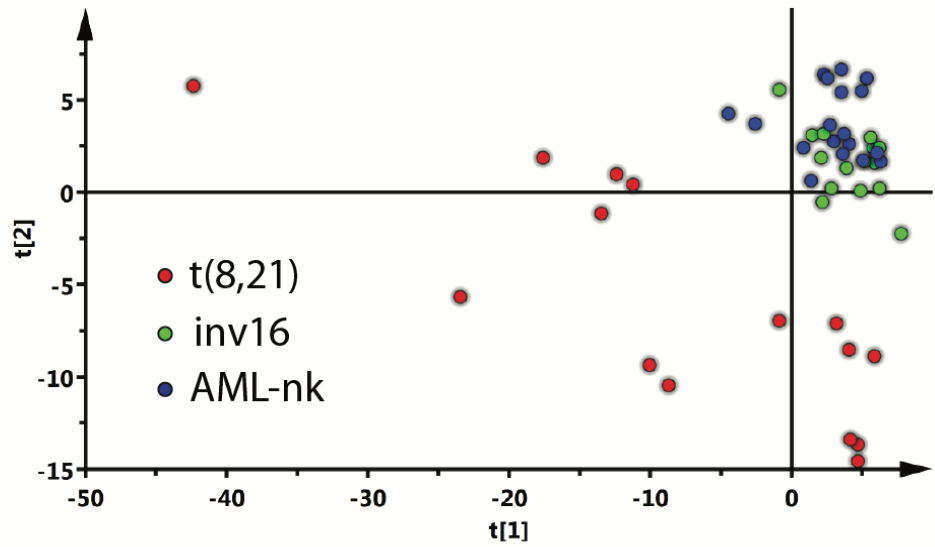

**Figure 1.** Score plots of the multivariate analysis by PCA. The t(8;21) samples are separated from the inv16 and AML-nk samples. Each patient sample was measured in triplicates and each point on the plot represents an individual measurement. The calculated R2X values are 0.253 and 0.094 for the first and second component, respectively. The analysis was performed using SIMCA 14.0 software.

Figure 2

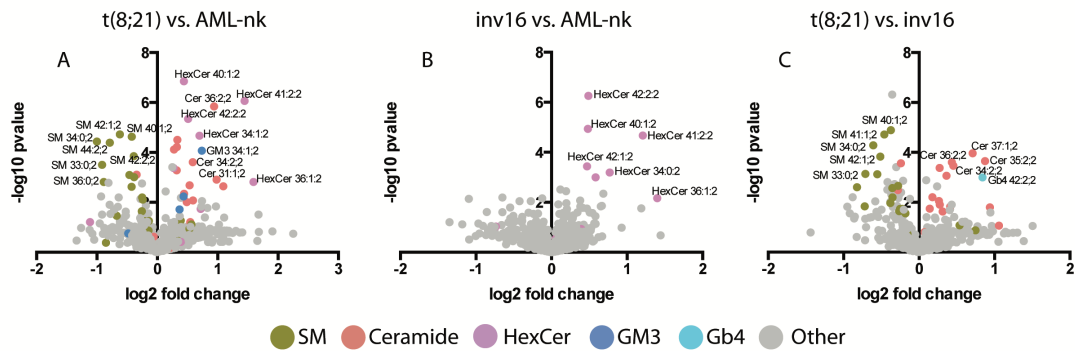

**Figure 2.** Lipidomic analysis of AML samples. Volcano plots were obtained from comparative analysis of 3 different AML types. Each point on the graph represents single lipid specie. The analysis was performed using SIMCA 14.0 software.

Figure 3

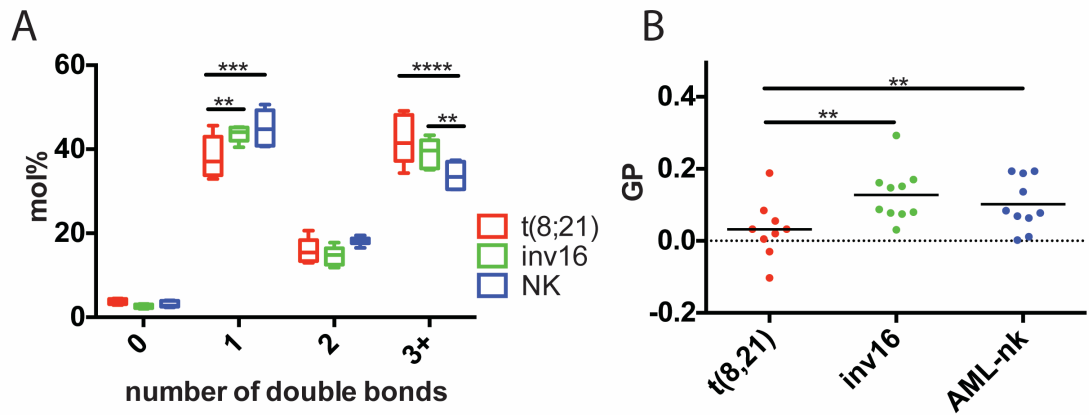

**Figure 3.** Lipid features extracted from AML samples. (A) Saturation index of all lipids measured with shotgun MS. The analysis revealed significant changes of the mono- and polyunsaturated fatty acids in t(8;21) samples. (B) GP index of liposomes prepared from lipid extracts of various AML samples. GP is a relative indicator of the membrane order (higher GP equals more ordered membranes). t(8;21) samples show lower GP values, which means that their membranes display higher fluidity.

Supplementary Figure 1

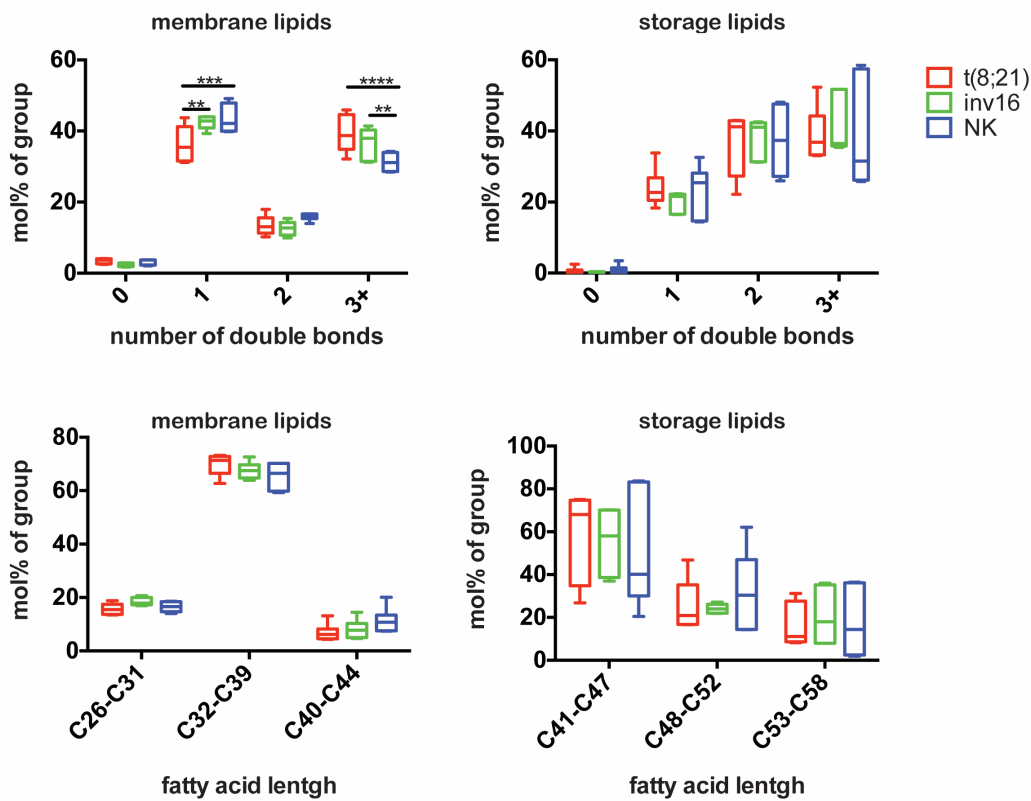

**Supplementary Figure 1** – Lipid features (fatty acid saturation and length) measured with shotgun MS of the membrane (PC, PC-O, PE, PE-O, PS, PI, PG, PA, SM, Cer, HexCer, DiHexCer, GM3, Gb3, Gb4) and storage lipids (SE, TAG, DAG).

**Table 1.** Description of patients.

| ID | karyotype | sex | age | FLT3-ITD mutation | FLT3-ITD ratio | NPM1 mutation | KIT mutation | first complete remission achieved | relapse after complete remission |
|---|---|---|---|---|---|---|---|---|---|
| 1 | normal | f | 81 | N | | Y | NA | N | N |
| 2 | normal | m | 59 | Y | 0.92 | Y | NA | Y | Y |
| 3 | normal | f | 62 | Y | 0.91 | Y | NA | N | N |
| 4 | normal | f | 72 | Y | 5.9 | Y | NA | N | N |
| 5 | normal | f | 72 | N | | Y | NA | Y | Y |
| 6 | normal | m | 63 | Y | NA | Y | NA | Y | N |
| 7 | inv16 | f | 53 | N | | N | N | N | N |
| 8 | inv16 | f | 30 | N | | N | N | Y | Y |
| 9 | inv16 | m | 28 | N | | N | Exon 8 | Y | Y |
| 10 | inv16 | m | 36 | N | | N | Exon 8 | Y | N |
| 11 | inv16 | m | 41 | N | | N | N | Y | Y |
| 12 | t(8;21) | m | 37 | N | | N | N | Y | N |
| 13 | t(8;21) | f | 26 | N | | N | D816V/H | Y | N |
| 14 | t(8;21) | m | 41 | N | | N | N | Y | N |
| 15 | t(8;21) | m | 53 | N | | N | D816V | Y | N |
| 16 | t(8;21) | f | 34 | N | | N | N | Y | Y |

f- female; m-male; NA-no information; N-no; Y-yes;